\documentclass{iaus}
\usepackage{graphicx}
\title[]
{Stellar Fluxes as Probes of Convection in Stellar Atmospheres}

\author[]{Barry Smalley}

\affiliation{Astrophysics Group, Keele University, Staffordshire ST5 5BG, United
Kingdom\break
email: bs@astro.keele.ac.uk}

\pubyear{2007}
\volume{IAUS239}
\pagerange{1--3}
\date{?? and in revised form ??}
\jname{Convection in Astrophysics}
\editors{F.Kupka, I.W. Roxburgh \& K.L. Chan, eds.}
\begin{document}

\maketitle

\begin{abstract}
Convection and turbulence in stellar atmospheres have a significant effect on the emergent flux from late-type stars. The theoretical
advancements in convection modelling over recent years have proved challenging for the observers to obtain measurements with sufficient
precision and accuracy to allow discrimination between the various predictions.

An overview of the current observational techniques used to evaluate various convection theories is presented, including photometry,
spectrophotometry, and spectroscopy. The results from these techniques are discussed, along with their successes and limitations.

The prospects for improved observations of stellar fluxes are also given.
\end{abstract}

\firstsection 
\section{Introduction}

The gross properties of a star, such as broad-band colours and flux
distributions, are significantly affected by the effects of convection in stars
later than mid A-type. Consequently, our modelling of convection in stellar
atmosphere models can significantly alter our interpretation of observed
phenomena. By comparison with stars of known $T_{\rm eff}$ and/or $\log g$ (the
fundamental stars), we can evaluate different treatments of convection in
model atmosphere calculations.

\section{Photometry}

Photometric indices are a fast and efficient method for determining approximate
atmospheric parameters of stars. For the commonly-used Str\"{o}mgren $uvby$
system a vast body of observational data exists which can be used to estimate
parameters using calibrated model grids (e.g. \cite[Moon \& Dworetsky
1985]{MD85}, \cite[Smalley \& Dworetsky 1995]{SD95}). Conversely, knowing
atmospheric parameters from other methods, allows observed colours to be
compared to model predictions. This method has been used to compare various
treatments of stellar convection.

The effects of convection on the theoretical $uvby$ colours of A, F, and G
stars was discussed by \cite{SK97}, who compared the predicted colours for the
\cite{CM91,CM92} (CM) model with that from the standard \cite{KUR93}
mixing-length theory (MLT) models with and without ``approximate
overshooting''. Comparison against fundamental $T_{\rm eff}$ and $\log g$ stars
revealed that the CM models gave better agreement than MLT without
overshooting. Models with overshooting were clearly discrepant. This result was
further supported by stars with $T_{\rm eff}$ obtained from the Infrared Flux
Method (IRFM) and $\log g$ from stellar evolutionary models.

\section{Fluxes}

\begin{figure}
\includegraphics{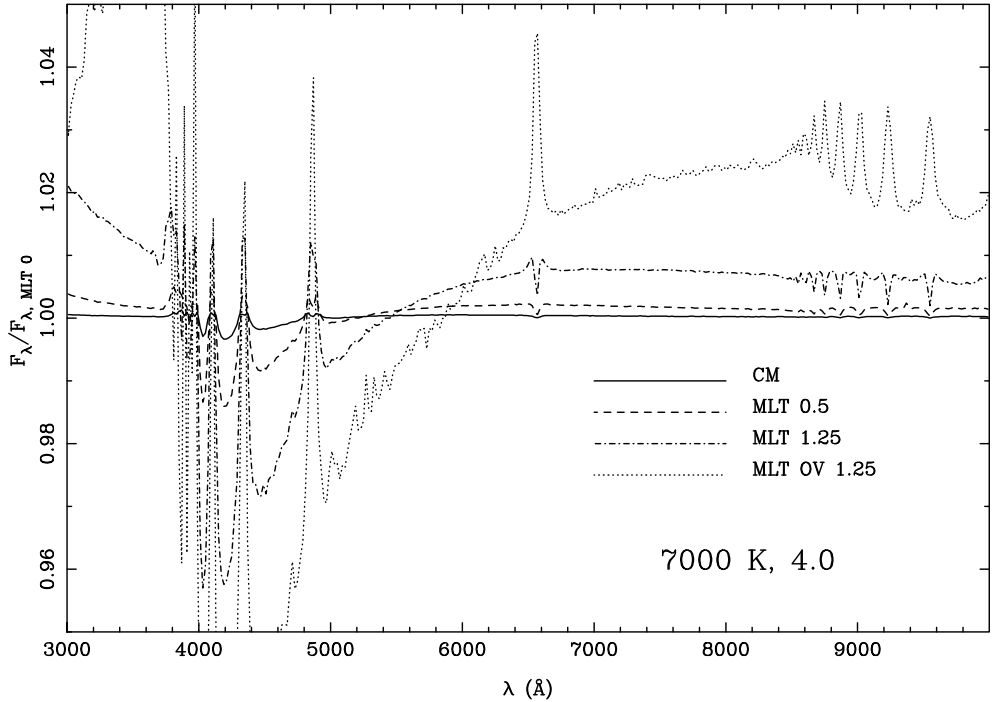}
\caption{Fluxes for solar-composition $T_{\rm eff}$ = 7000K, $\log g$ = 4
models with CM and MLT ($l/H$ = 0.5 and 1.25), compared to that for a model with
zero convection. Note that the region 4000 $\sim$ 5000\AA\ is especially sensitive and
the effect of overshooting is considerable.}

\label{smalley-fig}
\end{figure}

The observed stellar flux distribution is influenced by the effects of
convection on the atmospheric structure of the star. As we have seen with
photometric colours, these effects have a clearly observable signature (see
Fig.~\ref{smalley-fig}). In their discussion of convection \cite{LLK82}
presented model stellar atmospheres using a modified mixing-length theory. They
found small, systematic differences in the optical fluxes. Their figures also
demonstrate that convection can have a measurable effect on stellar fluxes.
Hence, high precision stellar flux measurements will provide significant and
useful information on convection.

Unfortunately, very little high-precision stellar spectrophotometry exists.
This situation will be rectified, once the ASTRA Spectrophotometer (see below) begins operation. This will allow spectrophotometry to
be added to our observational diagnostic toolkit.

\section{Balmer Profiles}

The temperature sensitivity of Balmer lines makes them an excellent diagnostic
tool for late A-type stars and cooler. The $H\alpha$ and $H\beta$ profiles
behave differently due to convection: $H\alpha$ is significantly less sensitive
to mixing-length than $H\beta$ (\cite[van't Veer \& M\'{e}gessier 1996]{VM96}).
Both profiles are affected by the presence of overshooting. Since $H\alpha$ is
formed higher in the atmosphere than $H\beta$, Balmer lines profiles are a very
good depth probe of stellar atmospheres. Balmer profiles are also
affected by microturbulence, metallicity and, for hotter stars, surface
gravity (\cite[Heiter \etal\ 2002]{HEI+02}).

In their comparison of Balmer line profiles, \cite{GKS99} found that both CM
and MLT without overshooting gave satisfactory agreement with fundamental
stars. Overshooting was again found to be discrepant. In addition, \cite{GKS99}
found evidence for significant disagreement between all treatments of
convection for stars with $T_{\rm eff}$ around 8000 $\sim$ 9000~K.
Subsequently, \cite{SMA+02} reviewed this region using binary systems with
known $\log g$ values and their revised fundamental $T_{\rm eff}$ values of the
component stars. They found that the discrepancy found was no longer as
evident. However, this region was relatively devoid of stars with fundamental
values of both $T_{\rm eff}$ and $\log g$. Further fundamental stars are
clearly required in this region.

\section{The ASTRA Spectrophotometer}

The Automated Spectrophotometric Telescope Research Associates (ASTRA) have
developed a cassegrain spectrophotometer and its automated 0.5-m f/16
telescope. There are being integrated at the Fairborn Observatory near Nogales,
Arizona.  Scientific observations are expected to begin in 2007 (\cite[Adelman
\etal\ 2007, Smalley \etal\ 2007]{ADE+07,SMA+07}).

In an hour the system will obtain S/N = 200 (after correction for instrumental
errors) observations of stars as faint as 9.5 mag.  The spectrograph uses both
a grating and a cross-dispersing prism to produce spectra from both the first
and the second orders simultaneously.  The square 30 arc second sky fields for
each order do not overlap.  The resolution is 7 \AA\ in second and 14 \AA\ in
first order. The wavelength range is of approximately
$\lambda\lambda$3300-9000.

\section{Conclusions}

The effects of convection on the stellar atmospheric structure can be
successfully probed using a variety of observational diagnostics (\cite[Smalley
2004]{SMA04}). The combination of photometric colours and Balmer-line profiles
has given us a valuable insight into the nature of convection in stars. High
quality observations that are currently available and those that will be in the
near future, will enable further refinements in our theoretical models of
convection and turbulence in stellar atmospheres.


\begin{thebibliography}{}
\bibitem[Adelman \etal\ (2007)]{ADE+07}
{Adelman S.J., Gulliver A.F., Smalley B., Pazder J.S., Younger P.F., Boyd L.J.,
Epand D., Younger T.} 2007, in: C. Sterken (ed.),
\textit{The Future of Photometric, Spectrophotometric and Polarimetric
Standardization}, (San Francisco: ASP), in press.

\bibitem[Canuto \& Mazzitelli (1991)]{CM91}
{Canuto V.M., Mazzitelli I.} 1991, \textit{ApJ} 370, 295

\bibitem[Canuto \& Mazzitelli (1992)]{CM92}
{Canuto V.M., Mazzitelli I.} 1992, \textit{ApJ} 389, 724

\bibitem[Gardiner \etal\ (1999)]{GKS99}
{Gardiner R.B., Kupka F., Smalley B.} 1999, \textit{A\&A} 347, 876

\bibitem[Heiter \etal\ (2002)]{HEI+02}
{Heiter U., Kupka F., van't Veer-Menneret C., Barban C., Weiss W.W., Goupil
M.-J., Schmidt W., Katz D., Garrido R.} 2002, \textit{A\&A} 392, 619

\bibitem[Kurucz (1993)]{KUR93}
{Kurucz R.L.} 1993, Kurucz CD-ROM 13: ATLAS9, SAO, Cambridge, USA

\bibitem[Lester \etal\ (1982)]{LLK82}
{Lester J.B., Lane M.C., Kurucz R.L.} 1982, \textit{ApJ} 260, 272

\bibitem[Moon \& Dworetsky (1985)]{MD85}
{Moon T.T., Dworetsky M.M.} 1985, \textit{MNRAS} 217, 305

\bibitem[Smalley (2004)]{SMA04}
{Smalley B.} 2004, in Zverko J., Ziznovsky J., Adelman S.J. \& Weiss W.W.
(eds.),
\textit{The A Star Puzzle}, Proc. IAU Symposium No. 224 (Cambridge University
Press), p.\ 131

\bibitem[Smalley \& Dworetsky (1995]{SD95}
Smalley B., Dworetsky M.M., 1995, \textit{A\&A} 293, 446

\bibitem[Smalley \& Kupka (1997)]{SK97}
Smalley B., Kupka F., 1997, \textit{A\&A} 328, 439

\bibitem[Smalley \etal\ (2002)]{SMA+02}
Smalley B., Gardiner R.B., Kupka F., Bessell M.S., 2002, \textit{A\&A} 395, 601

\bibitem[Smalley \etal\ (2007)]{SMA+07}
{Smalley B., Gulliver A.F., Adelman S.J.} 2007, in: C. Sterken (ed.),
\textit{The Future of Photometric, Spectrophotometric and Polarimetric
Standardization}, (San Francisco: ASP), in press.

\bibitem[van't Veer \& M\'{e}gessier (1996)]{VM96}
van't Veer C., M\'{e}gessier C., 1996, \textit{A\&A} 309, 879

\end{thebibliography}
\end{document}